\documentclass[11pt,twoside]{article}

\usepackage{asp2014}

\aspSuppressVolSlug
\resetcounters

\begin{document}

\title{RCSEDv2: homogenization of multi-wavelength photometric data.}

\author{Victoria Toptun,$^{1,2}$ Igor Chilingarian,$^3$ Ivan Katkov,$^4$ Kirill Grishin,$^{5,1}$ Anastasia Kasparova,$^1$ Sviatoslav Borisov,$^{1,6}$ Evgenii Rubtsov,$^1$ Vladimir Goradzhanov,$^{1,3}$ and Vladislav Klochkov$^3$}
\affil{$^1$Sternberg Astronomical Institute, M.V.~Lomonosov Moscow State University, Moscow, Russia}
\affil{$^2$Department of Physics, M.V. Lomonosov Moscow State University, Moscow, Russia}
\affil{$^3$Center for Astrophysics - Harvard and Smithsonian, Cambridge, USA}
\affil{$^4$Center for Astro, Particle, and Planetary Physics, NYU Abu Dhabi, Abu Dhabi, UAE}
\affil{$^5$Universit\'e de Paris, CNRS, Astroparticule et Cosmologie, F-75013 Paris, France}
\affil{$^6$Department of Astronomy, University of Geneva, Switzerland}

\paperauthor{Victoria Toptun}{victoria.toptun@voxastro.org}{0000-0003-3599-3877}{Sternberg Astronomical Institute, Lomonosov Moscow State University}{}{Moscow}{}{119234}{Russia}
\paperauthor{Igor Chilingarian}{igor.chilingarian@cfa.harvard.edu}{0000-0002-7924-3253}{Center for Astrophysics - Harvard and Smithsonian}{}{Cambridge}{}{02138}{USA}
\paperauthor{Ivan Katkov}{katkov.ivan@gmail.com}{0000-0002-6425-6879}{NYU Abu Dhabi}{Center for Astro, Particle, and Planetary Physics}{Abu Dhabi}{}{129188}{UAE}
\paperauthor{Anastasia Kasparova}{anastasya.kasparova@gmail.com}{0000-0002-1091-5146}{Sternberg Astronomical Institute, Lomonosov Moscow State University}{}{Moscow}{}{119234}{Russia}
\paperauthor{Sviatoslav Borisov}{sb.borisov@voxastro.org}{0000-0002-2516-9000}{University of Geneva}{Department of Astronomy}{Geneva}{}{}{Switzerland}
\paperauthor{Kirill Grishin}{kirillg6@gmail.com}{0000-0003-3255-7340}{Sternberg Astronomical Institute, Lomonosov Moscow State University}{}{Moscow}{}{119234}{Russia}
\paperauthor{Evgenii Rubtsov}{rubtsov602@gmail.com}{0000-0001-8427-0240}{Sternberg Astronomical Institute, Lomonosov Moscow State University}{}{Moscow}{}{119234}{Russia}
\paperauthor{Vladimir Goradzhanov}{goradzhanov.vs17@physics.msu.ru}{0000-0002-2550-2520}{Sternberg Astronomical Institute, Lomonosov Moscow State University}{}{Moscow}{}{119234}{Russia}
\paperauthor{Vladislav Klochkov}{vladislavk4481@gmail.com}{0000-0003-3095-8933}{M.V. Lomonosov Moscow State University}{Department of Physics}{Moscow}{}{119991}{Russia}



\begin{abstract}
RCSEDv2 (https://rcsed2.voxastro.org/), the second Reference Catalog of Spectral Energy Distributions of galaxies includes the largest homogeneously processed photometric dataset for 4 million galaxies assembled from several wide-field surveys. Here we describe the methodology of the photometric data homogenization. We first correct all photometric measurements for the foreground Galactic extinction, then convert them into the photometric system we adopted as a standard (GALEX + SDSS + UKIDSS + WISE). We computed aperture corrections into several pre-defined apertures by using published galaxy sizes / light profiles and image quality for each of the surveys. We accounted for k-corrections using our own analytic approximations. Such a homogeneous photometric catalog allows us to build fully calibrated SEDs for the galaxies in our sample (defined by the availability of their spectra) and enables direct scientific analysis of this unique extragalactic dataset. 
\end{abstract}



\section{Introduction and motivation}

Over the past decades, significant progress has been achieved in the quality and depth of wide-field imaging sky surveys in virtually all wavelength domain. However, they use different instruments, photometric systems, and methods of data processing. Therefore, for many scientific applications (e.g. to search a group of objects based on their specific features or compare physical properties of different objects), one needs to homogenize these datasets. A few years ago we implemented this approach in the Reference Catalog of Spectral Energy Distributions of galaxies (RCSED; \citealp{RCSED}). It contains the results of analysis of 800k spectra of non-active galaxies from SDSS DR7 \citep{SDSS_DR7} fitted against several grids of stellar population models using the {\sc NBursts} code \citep{2007MNRAS.376.1033C,2007IAUS..241..175C} complemented with photometric data from GALEX \citep{2005ApJ...619L...1M}, SDSS, and UKIDSS \citep{2007MNRAS.379.1599L}. Exploring this homogeneous dataset led to the discovery of statistically significant samples of rare objects, such as compact elliptical galaxies \citep{CZ15}, extended dwarf post-starburst galaxies \citep{sh2021NatAs.tmp..208G}, and active galactic nuclei powered by intermediate-mass black holes \citep{Chilingarian+18}.

Currently, we are working on the second version of the catalog, RCSEDv2, which contains spectroscopic and multiwavelength photometric data for 4.7 million galaxies out to intermediate redshifts ($z<1.2$). In this paper we present the photometric dataset of RCSEDv2 and the approaches we took to homogenize data from different surveys.

\section{RCSEDv2 photometric datasets and their homogenization}

\paragraph{Data sources}
The RCSEDv2 photometric catalog was assembled from the following sky surveys: GALEX (near-UV), SDSS, DESI Legacy Surveys \citep{2019AJ....157..168D}, Dark Energy Survey \citep{2021ApJS..255...20A}, PanSTARRS PS1 \citep{2020ApJS..251....7F}, VST ATLAS \citep{2015MNRAS.451.4238S}, SkyMapper DR2 \citep{2019PASA...36...33O}, KIDS \citep{2017A&A...604A.134D} (optical), UKIDSS, UHS \citep{2018MNRAS.473.5113D}, VHS \citep{2012A&A...548A.119C}, VIKING \citep{2013Msngr.154...32E} (NIR), CatWISE \citep{2020ApJS..247...69E} and unWISE \citep{2019ApJS..240...30S} (NIR/mid-IR). The datasets were retrieved using publicly available Virtual Observatory access services as well as GALEX CasJobs, WFCAM Science Archive, NOAO Data Lab. The compiled photometric catalog contains integrated photometry and aperture magnitudes matching spectral apertures for almost all galaxies included in RCSEDv2. 

\begin{figure}
\centering
\vskip -0.5cm
\includegraphics[width=0.77\hsize]{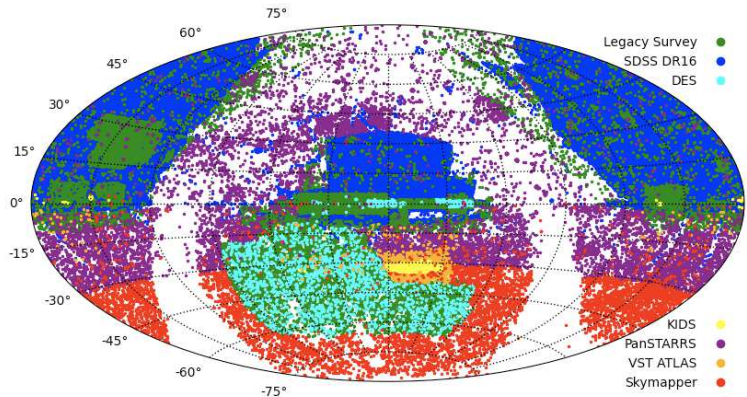}
 \\
\vspace{-0.5cm}
\caption{Sky coverage of optical photometric surveys included in RCSEDv2.
\label{cover}}
\vspace{-0.5cm}
\end{figure}

\paragraph{Extinction correction}
The first step is to correct photometric measurements for the Galactic foreground extinction. We used the e(B-V) maps and selective extinction coefficients from \citet{2011ApJ...737..103S}.

\paragraph{Conversion into spectral apertures}
To obtain photometric measurements consistent with the spectra from the spectroscopic datasets included in RCSEDv2, we converted aperture magnitudes from every photometric survey into the circular apertures with the diameters of 3'' (SDSS), 2'' (SDSS-BOSS and AAT-AAOmega surveys), 6.7'' (6dFGS), 1.5'' (Hectospec public spectra), 3.3'' (LAMOST). The conversion included interpolation and aperture corrections when needed. The aperture corrections were performed using 2D light profile models from SDSS for the objects from GALEX and WISE where the spatial resolution (6'') is worse than the aperture size.

\paragraph{Conversion into the ``standard'' photometric system}
The next step is to convert all magnitudes to some ``standard'' photometric system. Despite the same name, e.g. `SDSS g', the actual filter throughput curves differ significantly among different surveys. Therefore, one needs to adopt one reference filter set and convert all available measurements into it. We adopted the filter sets from GALEX, SDSS, UKIDSS, and WISE as internal reference and used the color transformation available from the literature to convert measurements from other surveys. In cases where no transformations were available, we derived them using data for spatially unresolved galaxies and excluding quasars, which always exhibit variability. One has to pay special attention to the fact that ``integrated'' photometry from different surveys may have different magnitude types, e.g. Petrosian in SDSS vs Kron in PanSTARRS. Also, Petrosian radii depend on the survey depth, therefore they are not the same for, e.g. SDSS and DES, making integrated photometry inconsistent. Therefore, one should  use integrated colors computed from photometric measurements originating from the same survey.

\begin{figure}
\centering
\vskip -4mm
\hfill
\includegraphics[width=0.3\hsize]{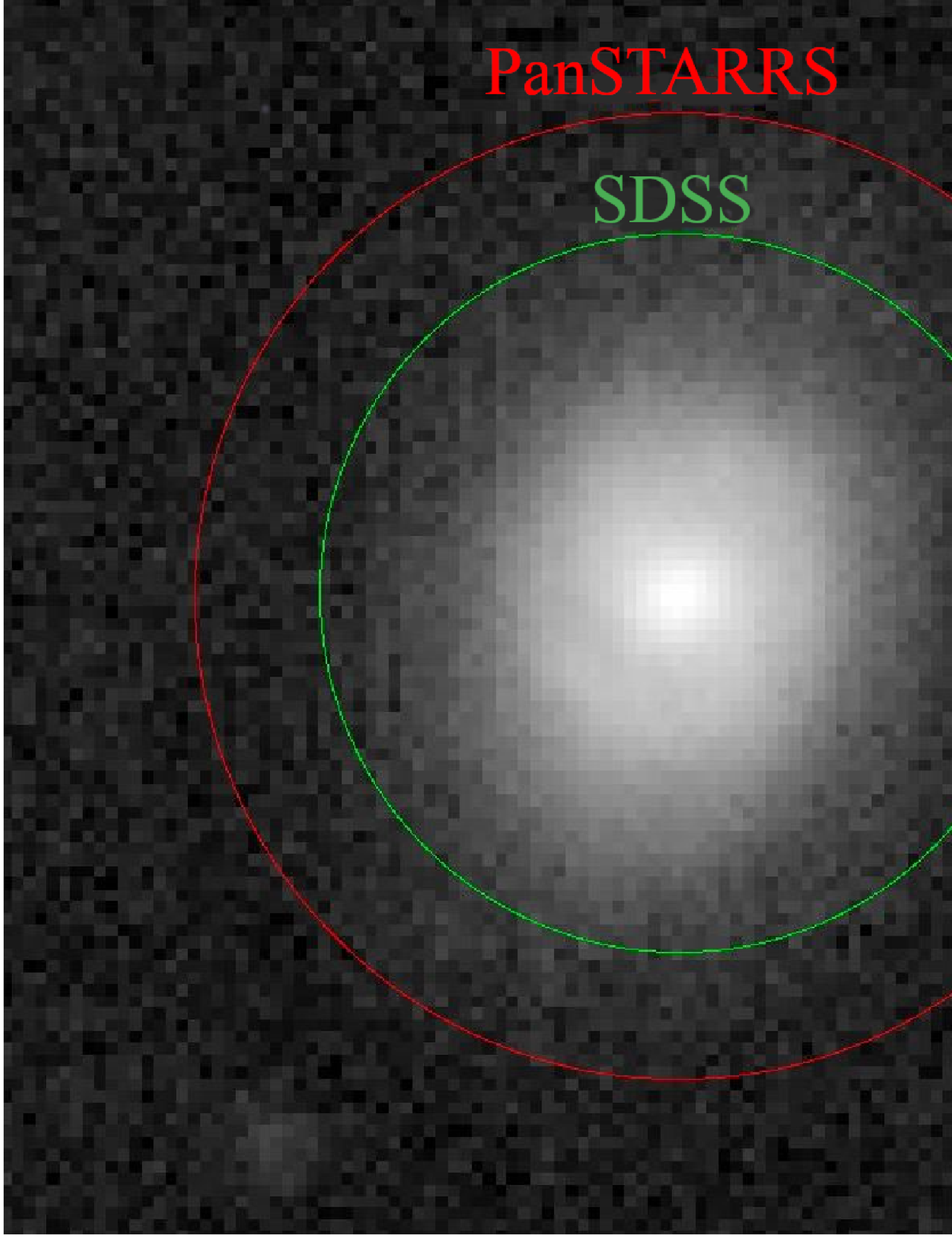}
\includegraphics[width=0.34\hsize]{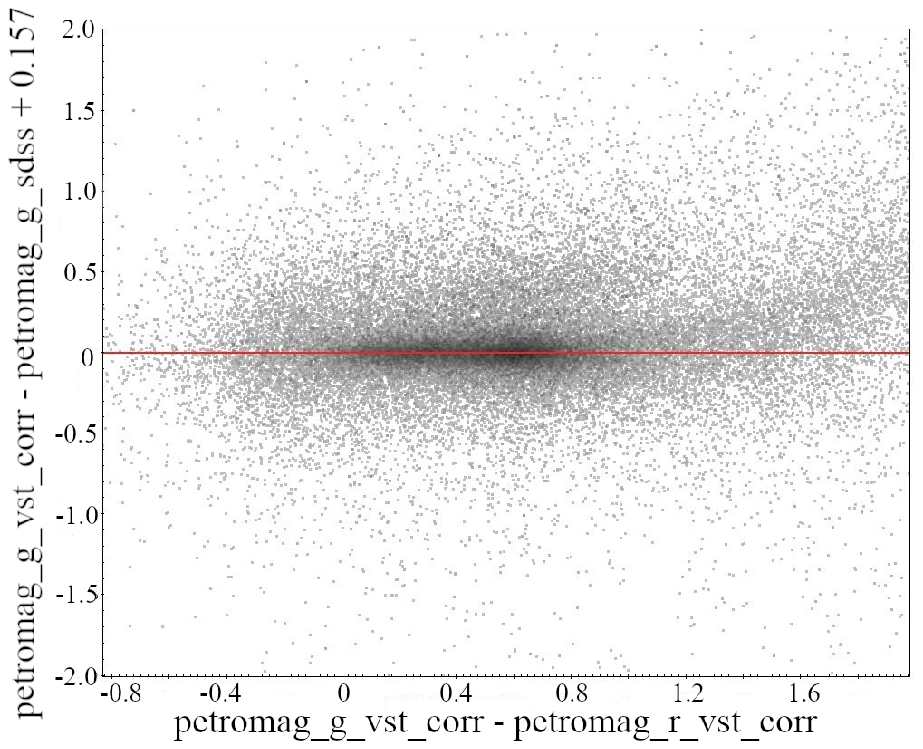}
\includegraphics[width=0.34\hsize]{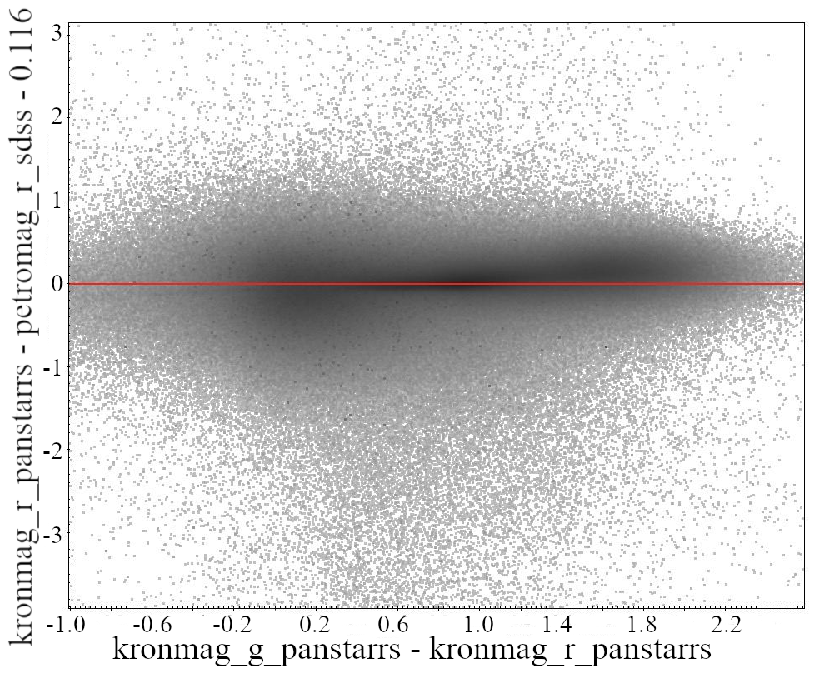} \\
\vspace{-0.4cm}
\caption{Left: Difference between the Petrosian radius of SDSS magnitude (green) and Kron radius of PanSTARRS (red); middle and right: the results of conversion of VST ATLAS and PanSTARRS into the SDSS photometric system
\label{conv}}
\vspace{-0.3cm}
\end{figure}

\paragraph{K-corrections}
To compare photometric properties of galaxies at different redshifts, we need to account for the shift of the filter throughput curve over the SED called $k$-correction. At $z>0.3$ the wavelength shift becomes so large that one should correct an observed magnitude in a given filter to a bluer restframe filter, e.g. observed $r$ into restframe $g$ (so called cross-band $k$-corrections). At low redshifts we use the analytic approximations of $k$-corrections available as kCorr\_*() functions in {\sc stilts} \citep{2006ASPC..351..666T} implementing polynomial approximations from \citet{2010MNRAS.405.1409C,2012MNRAS.419.1727C}. The approximations of cross-band $k$-corrections are described in a companion paper (Kasparova et al., this volume).

\section{RCSEDv2 database structure}
All RCSEDv2 are stored in the PostgreSQL 12.4 relational database management system with the pgSphere extension to handle spherical data types \citep{2004ASPC..314..225C}. Every galaxy has an association with all spectra collected within its $B$-band 25-th mag~arcsec$^{-2}$ elliptical isophote according to the sizes published in Hyperleda, Legacy Survey, SDSS, or VST ATLAS. Tables are linked using \emph{foreign key} constraints for consistency and indexed for fast data access.

\begin{figure}
\centering
\vskip -5mm
\includegraphics[width=0.55\hsize]{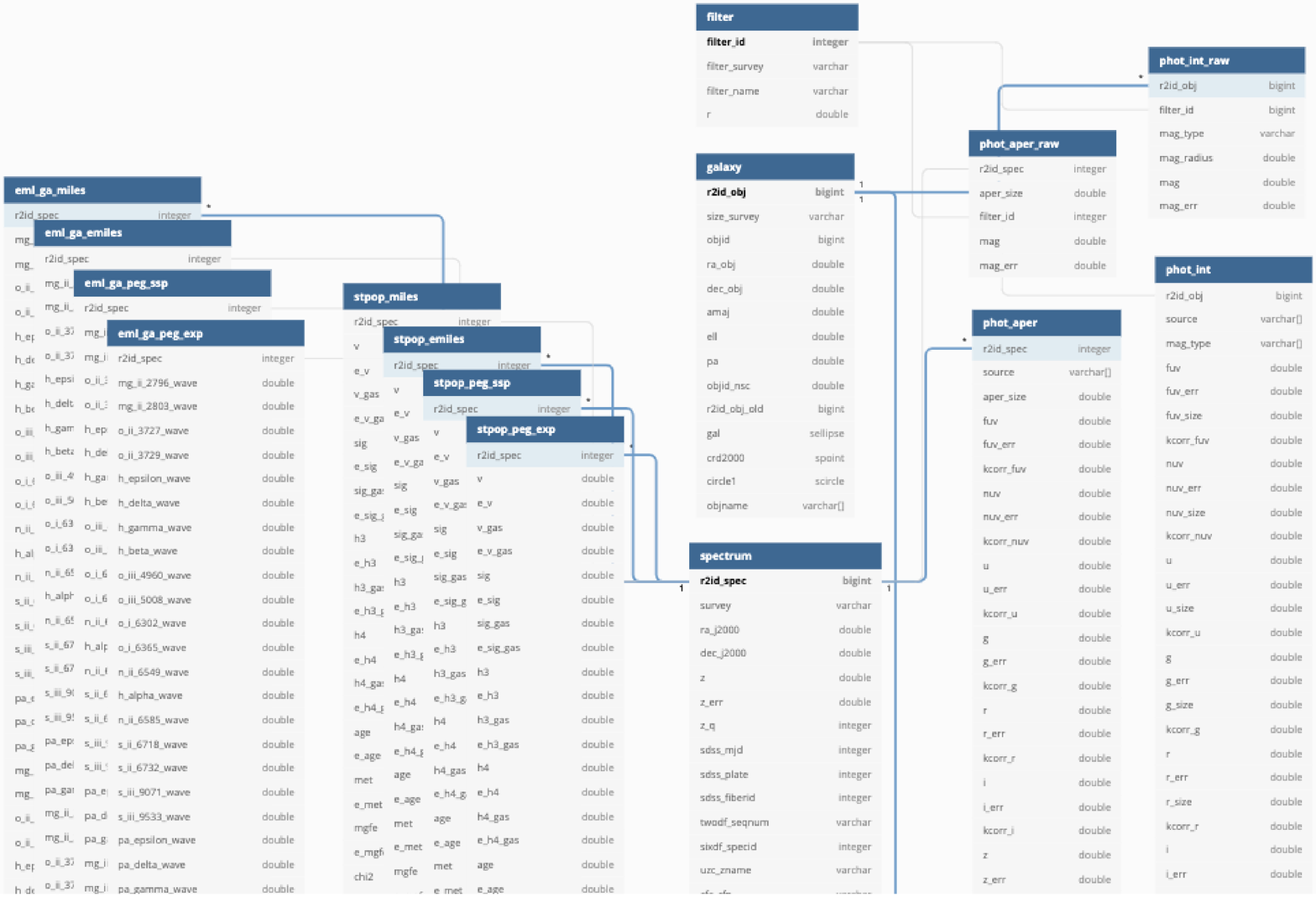}
 \\
\vspace{-0.3cm}
\caption{The RCSEDv2 relational database structure.
\label{db}}
\end{figure}
\vspace{-0.2cm}


\acknowledgements This project is supported by the RScF Grant 19-12-00281 the Interdisciplinary Scientific and Educational School of Moscow University ``Fundamental and Applied Space Research''. VT is grateful to the ADASS-XXXI organizing committee for providing financial aid to support her attendance of the conference.

\bibliography{O7-009}

\begin{thebibliography}{}
\expandafter\ifx\csname natexlab\endcsname\relax\def\natexlab#1{#1}\fi
\expandafter\ifx\csname url\endcsname\relax
  \def\url#1{\texttt{#1}}\fi
\expandafter\ifx\csname urlprefix\endcsname\relax\def\urlprefix{URL }\fi
\providecommand{\eprint}[2][]{\url{#2}}

\bibitem[{{Abazajian} et~al.(2009){Abazajian}, {Adelman-McCarthy},
  {Ag{\"u}eros}, {Allam}, {Allende Prieto}, {An}, {Anderson}, {Anderson},
  {Annis}, {Bahcall}, \& et~al.}]{SDSS_DR7}
{Abazajian}, K.~N., {et~al.} 2009, \apjs, 182, 543-558. \eprint{0812.0649}

\bibitem[{{Abbott} et~al.(2021){Abbott}, {Adam{\'o}w}, {Aguena}, {Allam},
  {Amon}, {Annis}, {Avila}, {Bacon}, {Banerji}, {Bechtol}, {Becker},
  {Bernstein}, {Bertin}, {Bhargava}, {Bridle}, {Brooks}, {Burke}, {Carnero
  Rosell}, {Carrasco Kind}, {Carretero}, {Castand er}, {Cawthon}, {Chang},
  {Choi}, {Conselice}, {Costanzi}, {Crocce}, {da Costa}, {Davis}, {De Vicente},
  {DeRose}, {Desai}, {Diehl}, {Dietrich}, {Drlica-Wagner}, {Eckert},
  {Elvin-Poole}, {Everett}, {Evrard}, {Ferrero}, {Fert{\'e}}, {Flaugher},
  {Fosalba}, {Friedel}, {Frieman}, {Garc{\'\i}a-Bellido}, {Gaztanaga},
  {Gelman}, {Gerdes}, {Giannantonio}, {Gill}, {Gruen}, {Gruendl}, {Gschwend},
  {Gutierrez}, {Hartley}, {Hinton}, {Hollowood}, {Honscheid}, {Huterer},
  {James}, {Jeltema}, {Johnson}, {Kent}, {Kron}, {Kuehn}, {Kuropatkin},
  {Lahav}, {Li}, {Lidman}, {Lin}, {MacCrann}, {Maia}, {Manning}, {Maloney},
  {March}, {Marshall}, {Martini}, {Melchior}, {Menanteau}, {Miquel}, {Morgan},
  {Myles}, {Neilsen}, {Ogand o}, {Palmese}, {Paz-Chinch{\'o}n}, {Petravick},
  {Pieres}, {Plazas}, {Pond}, {Rodriguez-Monroy}, {Romer}, {Roodman}, {Rykoff},
  {Sako}, {Sanchez}, {Santiago}, {Scarpine}, {Serrano}, {Sevilla-Noarbe},
  {Smith}, {Smith}, {Soares-Santos}, {Suchyta}, {Swanson}, {Tarle}, {Thomas},
  {To}, {Tremblay}, {Troxel}, {Tucker}, {Turner}, {Varga}, {Walker},
  {Wechsler}, {Weller}, {Wester}, {Wilkinson}, {Yanny}, {Zhang}, {Nikutta},
  {Fitzpatrick}, {Jacques}, {Scott}, {Olsen}, {Huang}, {Herrera}, {Juneau},
  {Nidever}, {Weaver}, {Adean}, {Correia}, {de Freitas}, {Freitas},
  {Singulani}, {Vila-Verde}, \& {Linea Science Server}}]{2021ApJS..255...20A}
{Abbott}, T.~M.~C., {et~al.} 2021, \apjs, 255, 20. \eprint{2101.05765}

\bibitem[{{Chilingarian} et~al.(2004){Chilingarian}, {Bartunov}, {Richter}, \&
  {Sigaev}}]{2004ASPC..314..225C}
{Chilingarian}, I., {et~al.} 2004, in Astronomical Data Analysis Software and
  Systems (ADASS) XIII, edited by F.~{Ochsenbein}, M.~G. {Allen}, \&
  D.~{Egret}, vol. 314 of Astronomical Society of the Pacific Conference
  Series, 225

\bibitem[{{Chilingarian} et~al.(2007{\natexlab{a}}){Chilingarian}, {Prugniel},
  {Sil'Chenko}, \& {Koleva}}]{2007IAUS..241..175C}
--- 2007{\natexlab{a}}, in Stellar Populations as Building Blocks of Galaxies,
  edited by A.~r. {Vazdekis}, \& R.~{Peletier}, vol. 241, 175.
  \eprint{0709.3047}

\bibitem[{{Chilingarian} \& {Zolotukhin}(2015)}]{CZ15}
{Chilingarian}, I., \& {Zolotukhin}, I. 2015, Science, 348, 418.
  \eprint{1504.06990}

\bibitem[{{Chilingarian} et~al.(2018){Chilingarian}, {Katkov}, {Zolotukhin},
  {Grishin}, {Beletsky}, {Boutsia}, \& {Osip}}]{Chilingarian+18}
{Chilingarian}, I.~V., {et~al.} 2018, \apj, 863, 1. \eprint{1805.01467}

\bibitem[{{Chilingarian} et~al.(2010){Chilingarian}, {Melchior}, \&
  {Zolotukhin}}]{2010MNRAS.405.1409C}
{Chilingarian}, I.~V., {Melchior}, A.-L., \& {Zolotukhin}, I.~Y. 2010, \mnras,
  405, 1409. \eprint{1002.2360}

\bibitem[{{Chilingarian} et~al.(2007{\natexlab{b}}){Chilingarian}, {Prugniel},
  {Sil'Chenko}, \& {Afanasiev}}]{2007MNRAS.376.1033C}
{Chilingarian}, I.~V., {et~al.} 2007{\natexlab{b}}, \mnras, 376, 1033.
  \eprint{astro-ph/0701842}

\bibitem[{{Chilingarian} \& {Zolotukhin}(2012)}]{2012MNRAS.419.1727C}
{Chilingarian}, I.~V., \& {Zolotukhin}, I.~Y. 2012, \mnras, 419, 1727.
  \eprint{1102.1159}

\bibitem[{{Chilingarian} et~al.(2017){Chilingarian}, {Zolotukhin}, {Katkov},
  {Melchior}, {Rubtsov}, \& {Grishin}}]{RCSED}
{Chilingarian}, I.~V., {et~al.} 2017, \apjs, 228, 14. \eprint{1612.02047}

\bibitem[{{Cross} et~al.(2012){Cross}, {Collins}, {Mann}, {Read}, {Sutorius},
  {Blake}, {Holliman}, {Hambly}, {Emerson}, {Lawrence}, \&
  {Noddle}}]{2012A&A...548A.119C}
{Cross}, N.~J.~G., {et~al.} 2012, \aap, 548, A119. \eprint{1210.2980}

\bibitem[{{de Jong} et~al.(2017){de Jong}, {Verdoes Kleijn}, {Erben},
  {Hildebrandt}, {Kuijken}, {Sikkema}, {Brescia}, {Bilicki}, {Napolitano},
  {Amaro}, {Begeman}, {Boxhoorn}, {Buddelmeijer}, {Cavuoti}, {Getman}, {Grado},
  {Helmich}, {Huang}, {Irisarri}, {La Barbera}, {Longo}, {McFarland},
  {Nakajima}, {Paolillo}, {Puddu}, {Radovich}, {Rifatto}, {Tortora},
  {Valentijn}, {Vellucci}, {Vriend}, {Amon}, {Blake}, {Choi}, {Conti}, {Gwyn},
  {Herbonnet}, {Heymans}, {Hoekstra}, {Klaes}, {Merten}, {Miller}, {Schneider},
  \& {Viola}}]{2017A&A...604A.134D}
{de Jong}, J. T.~A., {et~al.} 2017, \aap, 604, A134. \eprint{1703.02991}

\bibitem[{{Dey} et~al.(2019){Dey}, {Schlegel}, {Lang}, {Blum}, {Burleigh},
  {Fan}, {Findlay}, {Finkbeiner}, {Herrera}, {Juneau}, {Landriau}, {Levi},
  {McGreer}, {Meisner}, {Myers}, {Moustakas}, {Nugent}, {Patej}, {Schlafly},
  {Walker}, {Valdes}, {Weaver}, {Y{\`e}che}, {Zou}, {Zhou}, {Abareshi},
  {Abbott}, {Abolfathi}, {Aguilera}, {Alam}, {Allen}, {Alvarez}, {Annis},
  {Ansarinejad}, {Aubert}, {Beechert}, {Bell}, {BenZvi}, {Beutler}, {Bielby},
  {Bolton}, {Brice{\~n}o}, {Buckley-Geer}, {Butler}, {Calamida}, {Carlberg},
  {Carter}, {Casas}, {Castander}, {Choi}, {Comparat}, {Cukanovaite}, {Delubac},
  {DeVries}, {Dey}, {Dhungana}, {Dickinson}, {Ding}, {Donaldson}, {Duan},
  {Duckworth}, {Eftekharzadeh}, {Eisenstein}, {Etourneau}, {Fagrelius},
  {Farihi}, {Fitzpatrick}, {Font-Ribera}, {Fulmer}, {G{\"a}nsicke},
  {Gaztanaga}, {George}, {Gerdes}, {Gontcho}, {Gorgoni}, {Green}, {Guy},
  {Harmer}, {Hernandez}, {Honscheid}, {Huang}, {James}, {Jannuzi}, {Jiang},
  {Joyce}, {Karcher}, {Karkar}, {Kehoe}, {Kneib}, {Kueter-Young}, {Lan},
  {Lauer}, {Le Guillou}, {Le Van Suu}, {Lee}, {Lesser}, {Perreault Levasseur},
  {Li}, {Mann}, {Marshall}, {Mart{\'\i}nez-V{\'a}zquez}, {Martini}, {du Mas des
  Bourboux}, {McManus}, {Meier}, {M{\'e}nard}, {Metcalfe},
  {Mu{\~n}oz-Guti{\'e}rrez}, {Najita}, {Napier}, {Narayan}, {Newman}, {Nie},
  {Nord}, {Norman}, {Olsen}, {Paat}, {Palanque-Delabrouille}, {Peng},
  {Poppett}, {Poremba}, {Prakash}, {Rabinowitz}, {Raichoor}, {Rezaie},
  {Robertson}, {Roe}, {Ross}, {Ross}, {Rudnick}, {Safonova}, {Saha},
  {S{\'a}nchez}, {Savary}, {Schweiker}, {Scott}, {Seo}, {Shan}, {Silva},
  {Slepian}, {Soto}, {Sprayberry}, {Staten}, {Stillman}, {Stupak}, {Summers},
  {Sien Tie}, {Tirado}, {Vargas-Maga{\~n}a}, {Vivas}, {Wechsler}, {Williams},
  {Yang}, {Yang}, {Yapici}, {Zaritsky}, {Zenteno}, {Zhang}, {Zhang}, {Zhou}, \&
  {Zhou}}]{2019AJ....157..168D}
{Dey}, A., {et~al.} 2019, \aj, 157, 168. \eprint{1804.08657}

\bibitem[{{Dye} et~al.(2018){Dye}, {Lawrence}, {Read}, {Fan}, {Kerr},
  {Varricatt}, {Furnell}, {Edge}, {Irwin}, {Hambly}, {Lucas}, {Almaini},
  {Chambers}, {Green}, {Hewett}, {Liu}, {McGreer}, {Best}, {Zhang}, {Sutorius},
  {Froebrich}, {Magnier}, {Hasinger}, {Lederer}, {Bold}, \&
  {Tedds}}]{2018MNRAS.473.5113D}
{Dye}, S., {et~al.} 2018, \mnras, 473, 5113. \eprint{1707.09975}

\bibitem[{{Edge} et~al.(2013){Edge}, {Sutherland}, {Kuijken}, {Driver},
  {McMahon}, {Eales}, \& {Emerson}}]{2013Msngr.154...32E}
{Edge}, A., {et~al.} 2013, The Messenger, 154, 32

\bibitem[{{Eisenhardt} et~al.(2020){Eisenhardt}, {Marocco}, {Fowler},
  {Meisner}, {Kirkpatrick}, {Garcia}, {Jarrett}, {Koontz}, {Marchese},
  {Stanford}, {Caselden}, {Cushing}, {Cutri}, {Faherty}, {Gelino}, {Gonzalez},
  {Mainzer}, {Mobasher}, {Schlegel}, {Stern}, {Teplitz}, \&
  {Wright}}]{2020ApJS..247...69E}
{Eisenhardt}, P. R.~M., {et~al.} 2020, \apjs, 247, 69. \eprint{1908.08902}

\bibitem[{{Flewelling} et~al.(2020){Flewelling}, {Magnier}, {Chambers},
  {Heasley}, {Holmberg}, {Huber}, {Sweeney}, {Waters}, {Calamida}, {Casertano},
  {Chen}, {Farrow}, {Hasinger}, {Henderson}, {Long}, {Metcalfe}, {Narayan},
  {Nieto-Santisteban}, {Norberg}, {Rest}, {Saglia}, {Szalay}, {Thakar},
  {Tonry}, {Valenti}, {Werner}, {White}, {Denneau}, {Draper}, {Hodapp},
  {Jedicke}, {Kaiser}, {Kudritzki}, {Price}, {Wainscoat}, {Chastel}, {McLean},
  {Postman}, \& {Shiao}}]{2020ApJS..251....7F}
{Flewelling}, H.~A., {et~al.} 2020, \apjs, 251, 7. \eprint{1612.05243}

\bibitem[{{Grishin} et~al.(2021){Grishin}, {Chilingarian}, {Afanasiev},
  {Fabricant}, {Katkov}, {Moran}, \& {Yagi}}]{sh2021NatAs.tmp..208G}
{Grishin}, K.~A., {et~al.} 2021, Nature Astronomy. \eprint{2111.01140}

\bibitem[{{Lawrence} et~al.(2007){Lawrence}, {Warren}, {Almaini}, {Edge},
  {Hambly}, {Jameson}, {Lucas}, {Casali}, {Adamson}, {Dye}, {Emerson},
  {Foucaud}, {Hewett}, {Hirst}, {Hodgkin}, {Irwin}, {Lodieu}, {McMahon},
  {Simpson}, {Smail}, {Mortlock}, \& {Folger}}]{2007MNRAS.379.1599L}
{Lawrence}, A., {et~al.} 2007, \mnras, 379, 1599. \eprint{astro-ph/0604426}

\bibitem[{{Martin} et~al.(2005){Martin}, {Fanson}, {Schiminovich}, {Morrissey},
  {Friedman}, {Barlow}, {Conrow}, {Grange}, {Jelinsky}, {Milliard}, {Siegmund},
  {Bianchi}, {Byun}, {Donas}, {Forster}, {Heckman}, {Lee}, {Madore}, {Malina},
  {Neff}, {Rich}, {Small}, {Surber}, {Szalay}, {Welsh}, \&
  {Wyder}}]{2005ApJ...619L...1M}
{Martin}, D.~C., {et~al.} 2005, \apjl, 619, L1. \eprint{astro-ph/0411302}

\bibitem[{{Onken} et~al.(2019){Onken}, {Wolf}, {Bessell}, {Chang}, {Da Costa},
  {Luvaul}, {Mackey}, {Schmidt}, \& {Shao}}]{2019PASA...36...33O}
{Onken}, C.~A., {et~al.} 2019, PASA, 36, e033. \eprint{2008.10359}

\bibitem[{{Schlafly} \& {Finkbeiner}(2011)}]{2011ApJ...737..103S}
{Schlafly}, E.~F., \& {Finkbeiner}, D.~P. 2011, \apj, 737, 103.
  \eprint{1012.4804}

\bibitem[{{Schlafly} et~al.(2019){Schlafly}, {Meisner}, \&
  {Green}}]{2019ApJS..240...30S}
{Schlafly}, E.~F., {Meisner}, A.~M., \& {Green}, G.~M. 2019, \apjs, 240, 30.
  \eprint{1901.03337}

\bibitem[{{Shanks} et~al.(2015){Shanks}, {Metcalfe}, {Chehade}, {Findlay},
  {Irwin}, {Gonzalez-Solares}, {Lewis}, {Yoldas}, {Mann}, {Read}, {Sutorius},
  \& {Voutsinas}}]{2015MNRAS.451.4238S}
{Shanks}, T., {et~al.} 2015, \mnras, 451, 4238. \eprint{1502.05432}

\bibitem[{{Taylor}(2006)}]{2006ASPC..351..666T}
{Taylor}, M.~B. 2006, in Astronomical Data Analysis Software and Systems XV,
  edited by C.~{Gabriel}, C.~{Arviset}, D.~{Ponz}, \& S.~{Enrique}, vol. 351 of
  Astronomical Society of the Pacific Conference Series, 666

\end{thebibliography}


\end{document}